\documentclass{article}
\usepackage[dvips]{graphicx}
\begin{document}
\begin{center}
{\Large Stable localized pulses and zigzag stripes in a two-dimensional
diffractive-diffusive Ginzburg-Landau equation}

\bigskip

Hidetsugu Sakaguchi\footnote{{\normalsize e-mail:
sakaguchi@asem.kyushu-u.ac.jp}}

Department of Applied Science for Electronics and Materials,\\[0pt]
Interdisciplinary Graduate School of Engineering Sciences, Kyushu
University, Kasuga 816-8580, Japan
\end{center}

\bigskip

\begin{center}
Boris A. Malomed\footnote{{\normalsize e-mail: malomed@eng.tau.ac.il}}

Department of Interdisciplinary Studies, Faculty of Engineering, Tel Aviv
University, Tel Aviv 69978, Israel

\newpage

Abstract
\end{center}

We introduce a model of a two-dimensional (2D) optical waveguide with Kerr
nonlinearity, linear and quintic losses, cubic gain, and temporal-domain
filtering. In the general case, temporal dispersion is also included,
although it is not necessary. The model provides for description of a
nonlinear planar waveguide incorporated into a closed optical cavity. It
takes the form of a 2D cubic-quintic Ginzburg-Landau equation with an
anisotropy of a novel type: the equation is diffractive in one direction,
and diffusive in the other. By means of systematic simulations, we
demonstrate that the model gives rise to \emph{stable} fully localized 2D
pulses, which are spatiotemporal ``light bullets'', existing due to the
simultaneous balances between diffraction, dispersion, and Kerr
nonlinearity, and between linear and quintic losses and cubic gain. A
stability region of the 2D pulses is identified in the system's parameter
space. Besides that, we also find that the model generates 1D patterns in
the form of simple localized stripes, which may be stable, or may exhibit an
instability transforming them into oblique stripes with zigzags. The
straight and oblique stripes may stably coexist with the 2D pulse, but not
with each other. \newline

\newpage

\section{Introduction and formulation of the model}

Equations of the Ginzburg-Landau (GL) type constitute a class of universal
models to describe pattern formation and spatiotemporal chaos in nonlinear
media combining dissipative and dispersive properties \cite{CH}. The GL
model of the cubic-quintic (CQ) type was first introduced by Petviashvili
and Sergeev \cite{PetSer} in the two-dimensional (2D) case, with an
objective to construct stable fully localized 2D objects. In several works,
stable localized pulses have been obtained in the cubic-quintic GL equation
in one dimension (1D) \cite{Thual}. 2D pulses in models of the GL type were
considered by Thual and Fauve and by Deissler and Brand \cite{Deissler}.
Further, Firth and Scroggie predicted localized pulses in a two-dimensional
GL equation with the nonlinearity corresponding to a saturable absorber in
an optical cavity \cite{Firth}. Localized 2D pulse was also found in the CQ
equations of the Swift-Hohenberg type \cite{SB}. Recently, a stable
localized spiral pulse has been found in the 2D GL equation with the CQ
nonlinearity \cite{Bucharest}.

In previous works \cite{PetSer}-\cite{Bucharest}, localized pulses were
sought for in isotropic 2D equations of the GL type. In this work, we aim to
introduce a specific 2D GL equation which governs spatiotemporal evolution
of light in a planar active (i.e., equipped with an intrinsic gain)
nonlinear optical waveguide, and find stable fully localized pulse solutions
to it. The spatiotemporal 2D GL equation for the optical system is naturally
anisotropic, as the temporal variable plays the role of one of the
coordinates, see below. The existence of stable fully localized pulse
solutions to this equation suggest a new experimental realization of ``light
bullets'', i.e., spatiotemporal optical pulses. Thus far, ``bullets'' were
experimentally observed only in conservative second-harmonic-generating
optical media in an (effectively) 2D geometry \cite{Wise}. We aim to propose
a medium in which truly robust anisotropic 2D ``bullets'' exist as a result
of stable balance not only between the Kerr nonlinearity, spatial
diffraction, and temporal dispersion, but also between various loss and gain
mechanisms.

A prototype model equation, which governs the evolution of a local amplitude
$u(z,x,\tau )$ of the electromagnetic field in a 2D waveguide, is
\begin{equation}
iu_{z}+\frac{1}{2}u_{xx}+\frac{1}{2}\left( \beta -iF\right) u_{\tau \tau
}+|u|^{2}u=i\gamma _{0}u.  \label{proto}
\end{equation}
Here, $z$ and $x$ are the propagation and transverse coordinates,
\begin{equation}
\tau \equiv t-z/V_{0}  \label{t}
\end{equation}
is the so-called reduced time, where $t$ is the physical time, and $V_{0}$
is the group velocity of the carrier wave. The cubic term and the one $%
\,\sim u_{xx}$ in Eq. (\ref{proto}) account for the self-focusing Kerr
nonlinearity and transverse diffraction of the light pulse, respectively.
The linear terms with the coefficients $\beta $ and $F$ take into regard
chromatic dispersion and optical \textit{filtering}, which is formally
tantamount to diffusion in the $\tau $-direction. The term on r.h.s. of Eq. 
(%
\ref{proto}) with $\gamma _{0}>0$ is the optical gain (amplification). The
gain is necessary to compensate losses introduced by the filtering. Note
that filtering may be induced either by optical filters directly inserted
into the waveguide, or by a choice of an amplification mechanism which
provides for a \textit{bandwidth-limited} gain.

Due to its meaning, the filtering coefficient $F$ in Eq. (\ref{proto}) is
always positive, while the dispersion coefficient $\beta $ may be negative
(corresponding to normal dispersion), positive (anomalous dispersion), or 
$0$
(if the carrier wave is launched at the \textit{zero-dispersion point} \cite
{Agr}). In fact, just the latter case, $\beta =0$, is the most fundamental
one, as it corresponds to the simplest model of the present type that can
give rise to 2D pulses, see below.

The diffraction coefficient (the one in front of the term $u_{xx}$ in Eq. (%
\ref{proto})) in optical media has no imaginary part, hence Eq. 
(\ref{proto}%
) is intrinsically anisotropic: it features only \emph{diffraction} in the 
$%
x $-direction, and, in the above-mentioned simplest case $\beta =0$, only an
effective \emph{diffusion }(alias filtering) in the $\tau $-direction, while
$z$ plays the role of the evolutional variable. Other 2D anisotropic models
of the GL type are known in superconductivity (see, e.g., Refs. 
\cite{super}%
), as well as in the general pattern-formation theory \cite{Rabin} - \cite
{Hoyle}. However, in models of superconducting layers anisotropy is very
different from that in the present model; it is usually represented by an
asymmetric diffusion operator in the GL equation proper, which is coupled to
an equation for the magnetic field. In the above-mentioned pattern-forming
models, anisotropy is also drastically different from what we have in Eq. (%
\ref{proto}); in those models, the anisotropy is induced through a relative
asymmetry of the diffusion operators in two coupled GL equations \cite{DB},
or by addition of extra terms which directly break the axial symmetry of the
equation \cite{HS2}, or if a higher-order equation of the Segel - Whitehead
- Newell type is considered \cite{Hoyle}. To the best of our knowledge, no
2D model in which the GL equation is diffractive in one direction and
diffusive (or mixed diffusive/diffractive) in the other has been introduced
before.

Stationary solutions to Eq. (\ref{proto}) are sought for as
\begin{equation}
u(z,x,\tau )=e^{ikz}U(x,\tau ),  \label{stationary}
\end{equation}
where the complex function $U(x,\tau )$ obeys an equation
\begin{equation}
-kU+\frac{1}{2}U_{xx}+\frac{1}{2}\left( \beta -iF\right) U_{\tau \tau
}+|U|^{2}U=i\gamma _{0}U.  \label{Uproto}
\end{equation}
In particular, 1D pulse solutions of Eq. (\ref{proto}) have the form $%
U(x,\tau )=U(\theta )$, where
\begin{equation}
\theta \equiv \tau -px,  \label{theta}
\end{equation}
with an arbitrary real constant $p$. In this case, the function $U$ obeys an
equation
\begin{equation}
-kU+\frac{1}{2}\left[ p^{2}+\left( \beta -iF\right) \right] \frac{d^{2}U}{%
d\theta ^{2}}+|U|^{2}U=i\gamma _{0}U,  \label{Uproto1}
\end{equation}
and is assumed to vanish at $\theta \rightarrow \pm \infty $.

Equation (\ref{Uproto1}) coincides with the stationary version of the usual
1D complex cubic GL equation, and at all values of $p$ it has a single
solitary-pulse solution of the form \cite{PS}
\begin{equation}
U_{\mathrm{SP}}(\tau )=A\,\left[ \mathrm{sech}\left( \eta \theta \right) %
\right] ^{1+i\mu },  \label{PS}
\end{equation}
where
\begin{eqnarray*}
\mu  &=&\left( 2F\right) ^{-1}\left[ \sqrt{9\left( \beta +p^{2}\right)
+4F^{2}}-3\left( \beta +p^{2}\right) \right] , \\
\eta ^{2} &=&2\gamma _{0}\left[ F-2\left( \beta +p^{2}\right) \mu \right]
^{-1}, \\
A^{2} &=&3\left( \eta /2F\right) ^{2}\left[ \left( \beta +p^{2}\right) 
+F^{2}%
\right] \left[ \sqrt{9\left( \beta +p^{2}\right) +4F^{2}}-3\left( \beta
+p^{2}\right) \right] .
\end{eqnarray*}
Thus, Eq. (\ref{Uproto}) has the solution in the form of a 1D stripe
oriented in any direction in the ($x,\tau $) plane, except for that
corresponding to $p=0$, i.e., parallel to the $\tau $ axis, see Eq. (\ref
{theta}). This suggests to look for a solution in the form of a 2D pulse
localized both in $x$ and in $\tau $, i.e., satisfying the boundary
conditions
\begin{equation}
U(|x|=\infty )=U(|\tau |=\infty )=0.  \label{bc}
\end{equation}

In the context of the 2D nonlinear optical waveguide, the 2D pulse (``light
bullet'') is traveling at the carrier-wave's group velocity $V_{0}$ in the
laboratory reference frame, see Eq. (\ref{t}), therefore the most physically
relevant case, in which a stable ``bullet'' may be observable, is that when
it is traveling in a closed ring. The latter situation may be realized if
the nonlinear planar waveguide is a part of a closed optical \textit{cavity}
(resonator) equipped with intrinsic gain (see, e.g., Refs. \cite
{cavity,saturable}), so that Eq. (\ref{proto}) is an average equation for
the cavity. A more exotic, but also more straightforward, possibility to
realize the closed ring is to roll the planar waveguide described by Eq. (%
\ref{proto}) into a cylindrical surface, so that the propagation coordinate 
$%
z$ will become a cyclic one.

However, both 1D and 2D pulses, that may be found as solutions of Eq. (\ref
{Uproto}), are unstable as a solution to Eq. (\ref{proto}), because the
presence of the linear gain in this equation makes the zero solution, which
is the background of the localized pulse, unstable. Thus, to obtain
physically meaningful pulses, it is necessary to modify the model so that to
make its zero solution stable. One possibility is to linearly couple the
cubic GL equation to an extra linear dissipative equation, following the
pattern of similar 1D models \cite{couple,exact}. Physically, the
accordingly modified model corresponds to a dual-core waveguide, in which
one (nonlinear) core is active (has the gain), while the other one features
only losses, the filtering and nonlinearity in it being negligible. Within
the framework of the 1D geometry, it has been demonstrated that, in a vast
parameter region of the dual-core system, the zero solution is stable and,
simultaneously, there are two stationary solitary pulses (they were found in
an exact analytical form \cite{exact}, following the pattern of the exact
solution (\ref{PS})). As a result, the pulse with the larger amplitude is
stable \cite{couple,exact}. It is quite feasible that stable 2D pulses may
also exist in a 2D version of the dual-core model.

Another possibility to produce stable pulses is to insert directly into the
waveguide, or into the above-mentioned closed cavity incorporating the
waveguide, a \textit{nonlinear saturable absorber} (NSA); note that optical
resonators including gain (usually provided by the Erbium dopant), nonlinear
waveguide, and NSA are well known (in the 1D case) in the context of
soliton-generating lasers, see Refs. \cite{saturable} and references
therein. In the presence of NSA, it is easy to select the system's
parameters so that, for a weak light signal, the absorption is stronger than
the linear gain, hence the zero solution is stable. However, as the
absorption saturates with the increase of the signal's intensity, the system
may feature nonlinear gain, opening way to the coexistence of the stable
zero solution with stable pulses.

As it is well known, the simplest model describing a combination of the
linear gain and NSA is based on the above-mentioned GL equation of the
cubic-quintic (CQ) type. In particular, the accordingly modified 2D equation
(\ref{proto}) takes the form
\begin{equation}
iu_{z}+\frac{1}{2}u_{xx}+\frac{1}{2}\left( \beta -iF\right) u_{\tau \tau
}+|u|^{2}u=-i\gamma _{0}u+i\gamma _{1}|u|^{2}u-i\gamma _{2}|u|^{4}u,
\label{CQ}
\end{equation}
where the linear and quintic dissipation coefficients $\gamma _{0}$ and $%
\gamma _{2}$, as well as the cubic-gain coefficient $\gamma _{1}$, are
positive. One may fix $\gamma _{0}\equiv 1$ and $F\equiv 1$ by means of
obvious rescalings, then the eventual form of Eq. (\ref{CQ}) is
\begin{equation}
iu_{z}+\frac{1}{2}u_{xx}+\frac{1}{2}\left( \beta -i\right) u_{\tau \tau }=-%
\left[ iu+\left( 1-i\gamma _{1}\right) |u|^{2}u+i\gamma _{2}|u|^{4}u\right] 
.
\label{CQfinal}
\end{equation}

Strictly speaking, in the case $\beta =0$ (at the zero-dispersion point),
which will be considered, among other cases, below, one should add the
third-order dispersion (TOD) to Eq. (\ref{CQfinal}), so that it will take
the form \cite{Agr}
\begin{equation}
iu_{z}+\frac{1}{2}u_{xx}+\frac{1}{2}\left( \beta -i\right) u_{\tau \tau }-%
\frac{1}{6}i\widetilde{\beta }u_{\tau \tau \tau }=-\left[ iu+\left(
1-i\gamma _{1}\right) |u|^{2}u+i\gamma _{2}|u|^{4}u\right] ,  \label{TOD}
\end{equation}
where $\widetilde{\beta }$ is a real TOD coefficient. However, TOD is
important only for pulses which are extremely narrow in the temporal
direction, therefore this extra term will not be taken into regard in most
cases in this work (but, nevertheless, the effect of this term on the 2D
pulses will be considered below). The model may also include a quintic
correction to the Kerr nonlinearity, but such a term it is not considered in
this work.

In the rest of the paper, an objective is to find, by means of direct
simulations of Eq. (\ref{CQfinal}), various stable stationary states in the
present model, both 2D solitary pulses (corresponding to the 
spatiotemporain{document}
``light bullets'' in the waveguide model formulated above) and (quasi) 1D
states. The latter ones will be represented by stripe patterns which are
localized in one direction, and by more sophisticated patterns that include
steps (they are similar to zigzag patterns known in some other models \cite
{HS,Hoyle}). Results for the 2D pulses are presented in section 2, and 1D
patterns are described in a brief form in section 3. Section 4 concludes the
paper.

\section{Stable two-dimensional localized pulses}

Equation (\ref{CQfinal}) was solved numerically by means of a pseudospectral
code with $256\times 256$ modes and periodic boundary conditions in $x$ and 
$%
\tau $, keeping fixed sizes of the system in both directions, $L_{x}=L_{\tau
}\equiv L=20$. To generate the first stable 2D pulse, an initial
configuration was taken in the form of a Gaussian, $u=2\exp \left[ -\left(
\left( x-L/2\right) ^{2}+\left( \tau-L/2\right) ^{2}\right) /4\right] $
(i.e., the center of the pulse is placed at $x=\tau =L/2$). In order to
generate stability diagrams (see below), we then varied the parameters by
small steps, the initial configuration for each simulation being the stable
pulse produced by the simulation at the previous step.
\begin{figure}[htb]
\begin{center}
\includegraphics[width=8cm]{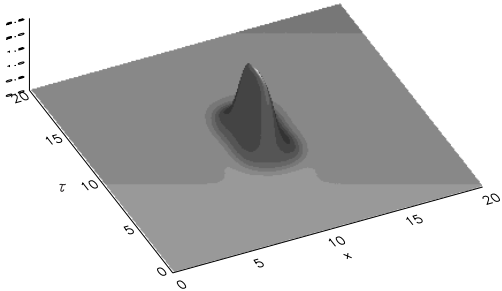}
\caption{A typical example of a stable stationary two-dimensional pulse found
in Eq. (\ref{CQfinal}) at $\beta =0$, $\gamma _{1}=2.5$, $\gamma _{2}=1/2$.
The absolute value of the field is shown as a function of $x$ and $\tau $.} 
\label{fig:1} 
\end{center}
\end{figure} 
\begin{figure}[htb]
\begin{center}
\includegraphics[width=11cm]{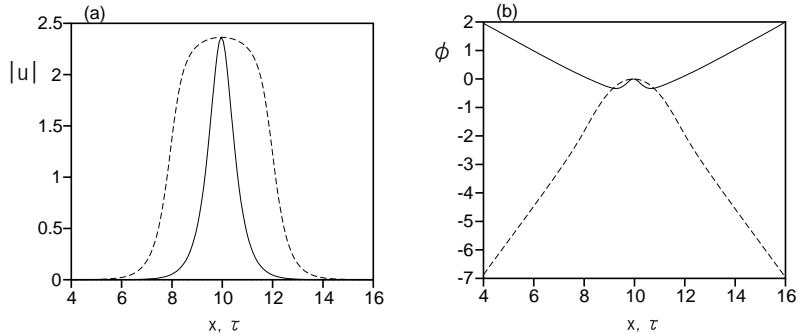}
\caption{The amplitude (a) and phase (b) profiles of the same stable soliton
which was shown in Fig. 1, taken along the cross sections $\tau =L/2$
(continuous lines) and $x=L/2$ (dashed lines).} 
\label{fig:2} 
\end{center}
\end{figure} 

A typical example of a stable fully localized pulse produced by the
simulations is displayed in Fig. 1, which pertains to the simplest and most
fundamental case, $\beta =0$. To further illustrate the structure of the
pulse, in Fig. 2 we display its field, represented in the form $u(z,x,\tau
)\equiv a(x,\tau )\exp \left[ ikz+i\phi (x,\tau )\right] $ with real
amplitude $a$ and phase $\phi $, along two principal cross sections, $x=L/2$
and $\tau =L/2$ . It is noteworthy that the pulse is much narrower in the 
$x$
(spatial) direction than in the temporal $\tau $-direction, which may be
explained as follows: the diffusion (optical filtering) in the $\tau $%
-direction, which is a dissipative feature, produces an essentially stronger
effect than diffraction (a conservative feature) along the $x$-direction;
hence, to compensate the nonlinear terms, a much larger curvature of the
soliton's profile at its center is necessary in the $x$-direction than in
the direction of $\tau $. It is noteworthy that the so-called chirp, i.e.,
curvature of the phase profiles, which plays an important role in dynamics
of solitons in nonlinear optics \cite{Agr}, is strongly concentrated at the
center of the pulse; note also a peculiar shape of the phase field in the 
$x$%
-direction.

\begin{figure}[htb]
\begin{center}
\includegraphics[width=11cm]{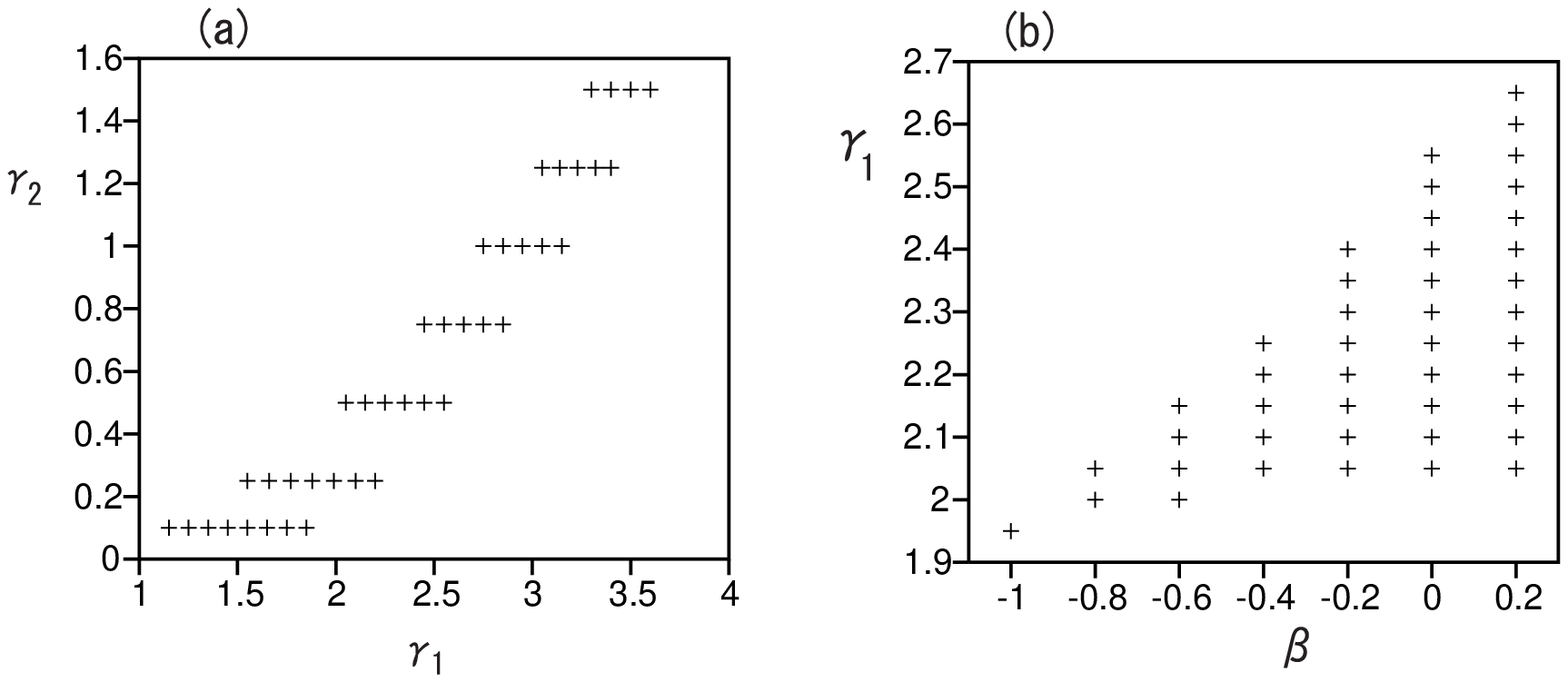}
\caption{Stability regions of the stationary two-dimensional solitary pulse
in the parametric planes ($\gamma _{1},\gamma _{2}$) for $\beta =0$ (a), and
($\beta ,\gamma _{1}$) for fixed $\gamma _{2}=1/2$ (b).} 
\label{fig:3} 
\end{center}
\end{figure} 
Results of many runs of the numerical simulations at different values of the
control parameters were collected in the form of stability diagrams
displayed in Fig. 3. They show regions in the parameter plane ($\gamma
_{1},\gamma _{2}$), at fixed $\beta =0$, and in the plane ($\beta ,\gamma
_{1}$), at fixed $\gamma _{2}=1/2$, in which \emph{stable} 2D pulses have
been found. The fact that the stability region expands while the dispersion
coefficient $\beta $ is changing from negative to positive, see Fig. 3(b),
can be easily understood: the existence of a solitary pulse as a result of
the balance between the self-focusing Kerr nonlinearity and anomalous
dispersion (corresponding to $\beta >0$) is much more natural than in the
case when the self-focusing nonlinearity competes with normal dispersion
(which corresponds to $\beta <0$) \cite{Agr}. Nevertheless, stable pulses
are found in the normal-dispersion area too. The pulse can exist in this
case because the effect of the diffusion (filtering) in the $\tau $%
-direction beats the anti-localization effect exerted by the normal
dispersion.

\begin{figure}[htb]
\begin{center}
\includegraphics[width=7cm]{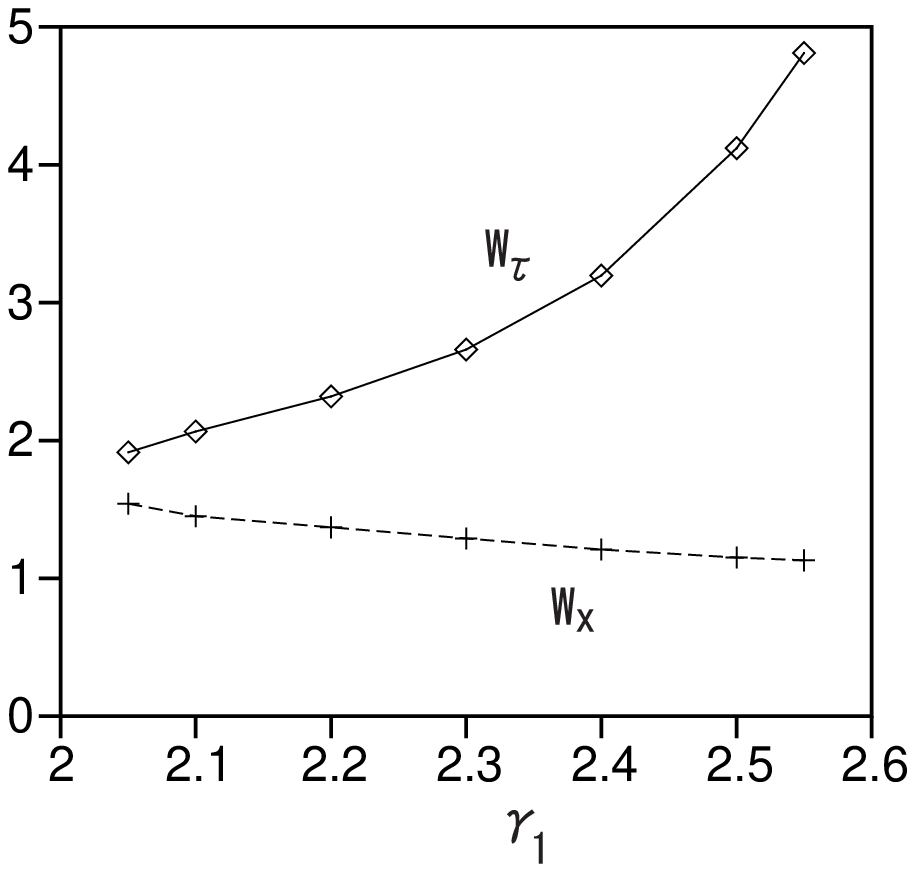}
\caption{The widths of the stable stationary two-dimensional pulse in the 
$x$%
-direction (rhombuses) and $\tau $-direction (crosses) vs. the cubic gain $%
\gamma _{1}$, at fixed values $\beta =0$ and $\gamma _{2}=1/2$.} 
\label{fig:4} 
\end{center}
\end{figure} 
To illustrate the change of the pulse's shape while varying parameters
inside the stability region, in Fig. 4 we display its widths $W_{\tau }$ and
$W_{x}$ of the pulse in the $\tau $- and $x$-directions as functions of $%
\gamma _{1}$. In fact, the results presented in Fig. 4 pertain to a cut
across the stability area delineated by Figs. 3(a) and 3(b). The cut is
taken in the direction parallel to the $\gamma _{1}$ axis, at fixed values 
$%
\gamma _{2}=1/2$ and $\beta =0$. The widths $W_{\tau }$ and $W_{x}$ were
measured along the cross sections $x=L/2$ and $\tau =L/2$ of the pulse, as
separation between points where, respectively, $|u(x=L/2,\tau \equiv L/2\pm
W_{\tau }/2)|=(1/2)|u|_{\max }$, and $|u(x\equiv L/2\pm W_{x}/2,\tau
=L/2)|=(1/2)|u|_{\max }$, with $|u|_{\max }$ taken at the center of the
pulse, $x=\tau =L/2$. As well as in the case shown in Figs. 1 and 2, one
concludes, looking at Fig. 4, that the pulse is essentially narrower in the 
$%
x$-direction.

The simulations demonstrate that, beyond the left border of the stability
region in Fig. 3(a) and beyond the lower border of the stability region in
Fig. 3(b), any initially created pulse decays to zero. This is quite
understandable, as in the case when the gain coefficient $\gamma _{1}$ is
too small, nothing but the trivial solution may exist as a steady state,
i.e., no stationary pulse is present in this case. On the other hand,
various transitions to spatially extended patterns take place beyond the
right border of the stability region in Fig. 3(a) and beyond the upper
boundary in Fig. 3(b). For example, a persistent spatially extended and
temporally chaotic state sets in for $\gamma _{1}>2.6$ at $\beta =0$ and $%
\gamma _{2}=1/2$. A transition to a localized zigzag (stepped) stripe state
(discussed in the next section) takes place for $\gamma _{1}>2.3$ at $\beta
=-0.4$ and $\gamma _{2}=1/2$.

\begin{figure}[htb]
\begin{center}
\includegraphics[width=8cm]{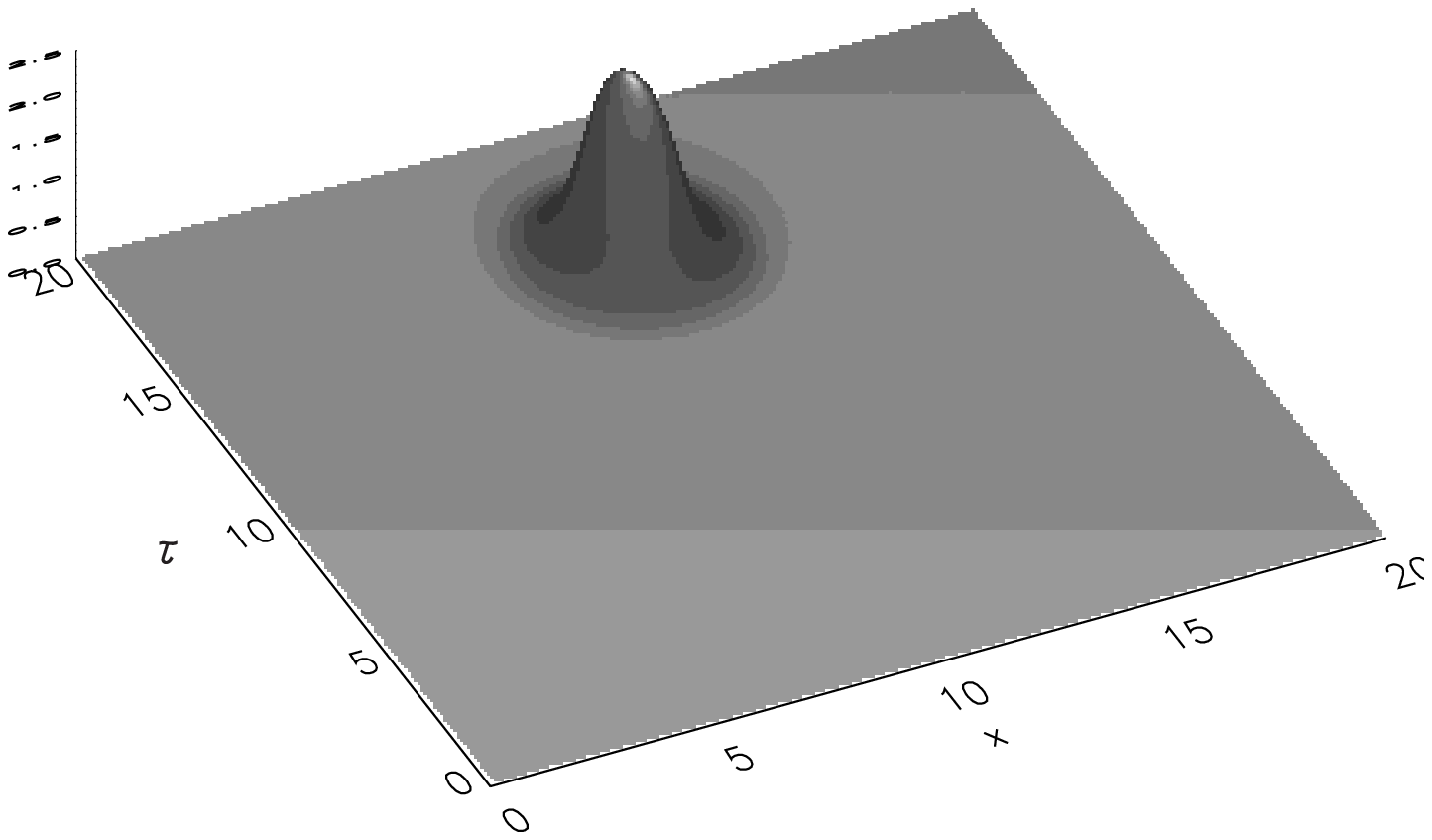}
\caption{A snapshot of a two-dimensional pulse obtained from Eq. (\ref{TOD}),
that includes the third-order-dispersion term with the coefficient $\tilde{%
\beta}=0.1$, the other parameters being the same as in Fig. 1. The absolute
value of the field is shown as a function of $x$ and $\tau $.} 
\label{fig:5} 
\end{center}
\end{figure} 
\begin{figure}[htb]
\begin{center}
\includegraphics[width=11cm]{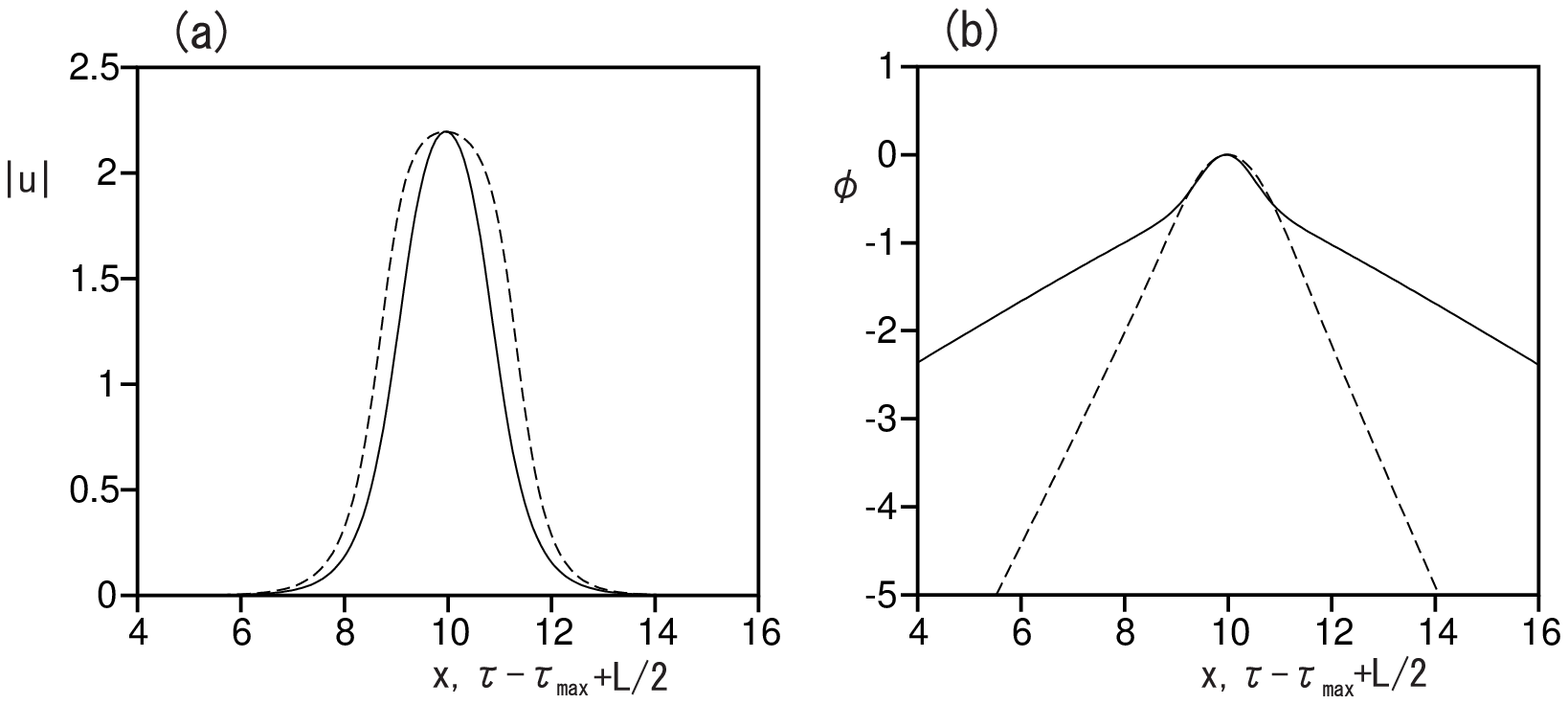}
\caption{The amplitude (a) and phase (b) profiles of the pulse shown in Fig.
5, taken along the cross sections $\tau =\tau _{\max }$ (continuous lines)
and $x=L/2$ (dashed lines). The $\tau $ axis is offset by $L/2-\tau _{\max 
}$
to stress that the peak positions of the amplitude and phase profiles
coincide.} 
\label{fig:6} 
\end{center}
\end{figure} 
To check the effect of the third-order dispersion (TOD), a few numerical
simulations of the accordingly modified equation (\ref{TOD}) were performed.
Figure 5 shows a snapshot of the 2D localized pulse affected by TOD with $%
\tilde{\beta}=0.1$. The pulse is more circular than the one shown for the
same values of the parameters, except that $\tilde{\beta}=0$, in Fig. 1, and
it is moving along the $\tau $ direction. Figures 6 (a) and (b) display the
amplitude profiles $a$ and the phase profiles $\phi $ along two principal
cross sections, $x=L/2$ and $\tau =\tau _{\max }$, at the intersection of
which the amplitude $a(x,\tau )$ has its maximum ($\tau _{\max }$ is not
fixed as the pulse is moving). The width of the amplitude profile along the 
$%
\tau $ direction is smaller than that for $\tilde{\beta}=0$ shown in Fig.
2(a), plausibly because the additional dispersion provided by TOD makes the
pulse structure sharper. The phase profile along the $x$ direction is
strongly different from that in Fig. 2(b). Note also that the phase profile
along the $\tau $ direction is slightly asymmetric, which is a natural
consequence of the symmetry breaking by the TOD term.

\section{One-dimensional localized pulses and the zigzag instability}

The 2D localized pulse is not the single stable pattern generated by Eq. (%
\ref{CQfinal}). Simulations also reveal a stable 1D pattern in the form of a
straight stripe localized in the $x$-direction and uniform along the $\tau $
axis. As it was mentioned in the previous section, this type of 1D pulses
can be found when the 2D pulse becomes unstable (but it can also stably
coexist with a 2D pulse, see below). In terms of the underlying
optical-waveguide model, this solution corresponds to a spatial soliton,
rather than a spatiotemporal one corresponding to the 2D pulse considered in
the previous section. Note that Eq. (\ref{CQfinal}) is invariant against a
Galilean-like transformation in the spatial domain with an arbitrary real
parameter $q$,
\begin{equation}
u(z,x,\tau )=\exp \left( iqx-\frac{1}{2}iq^{2}z\right) \,\widetilde{u}\left(
z,\widetilde{x},\tau \right) ,\,\widetilde{x}\equiv x-qz,  \label{Galileo}
\end{equation}
which implies a possibility of an arbitrary orientation of the spatial
stripe soliton in the ($z,x$) plane.

Obviously, there also exist more general steady-state solutions to Eq. (\ref
{CQfinal}) in the form of a stripe oriented at oblique directions in the
spatiotemporal ($x,\tau $) plane, similar to the solution (\ref{PS}) to Eq. 
(%
\ref{proto}). In the optical waveguide, they correspond to moving spatial
solitons. These oblique stripe patterns will appear below.

\begin{figure}[htb]
\begin{center}
\includegraphics[width=8cm]{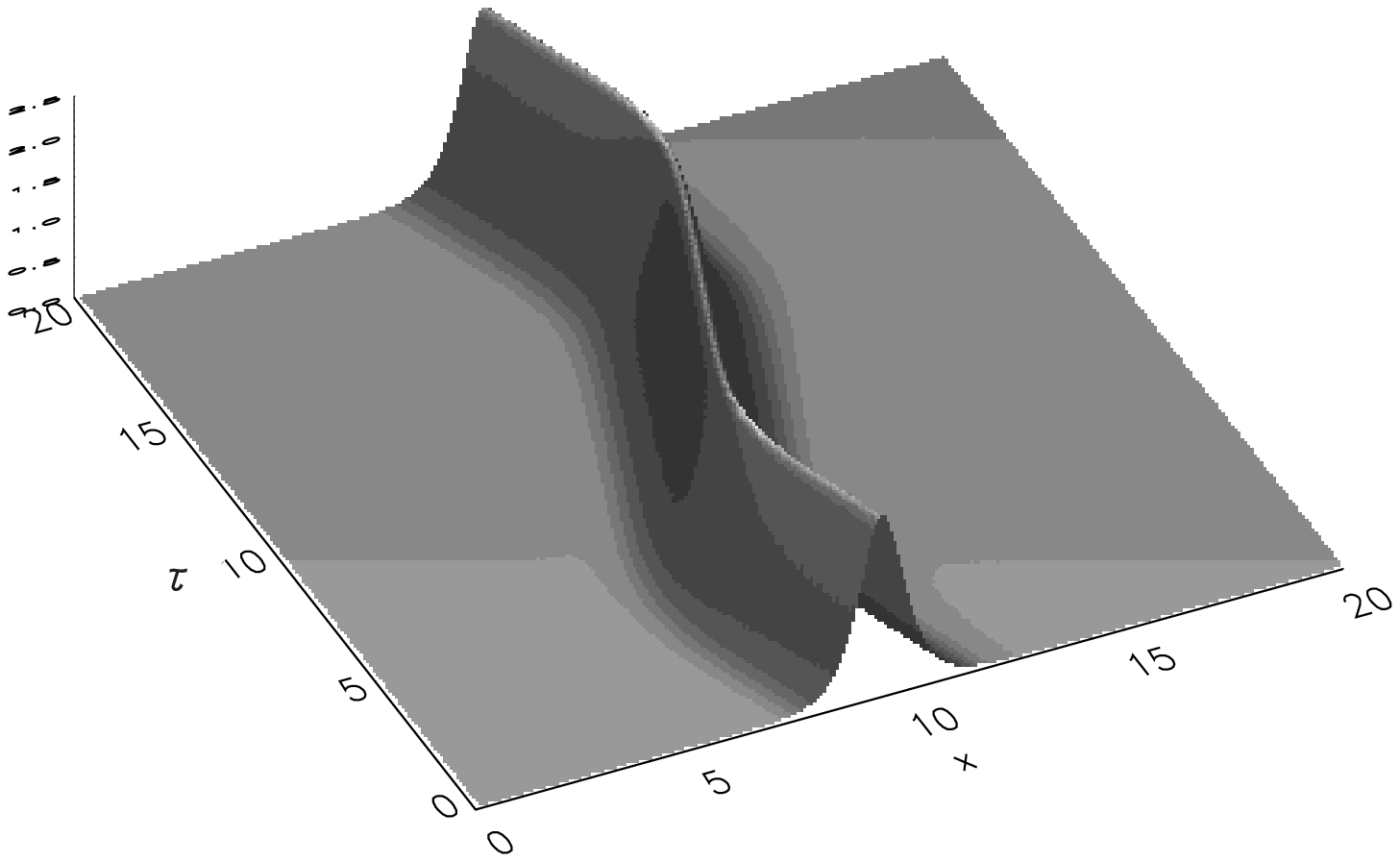}
\caption{An example of a quasi-one-dimensional stable stationary pattern
containing a zigzag. The absolute value of the field is shown as a function
of $x$ and $\tau $. This solution to Eq. (\ref{CQfinal}) was found at $\beta
=-0.05$, $\gamma _{1}=2.3$, $\gamma _{2}=1/2$.} 
\label{fig:7} 
\end{center}
\end{figure} 
Extended simulations demonstrate that the spatial solitons (stationary
stripes) and 2D spatiotemporal pulses coexist, as stable solutions to Eq. (%
\ref{CQfinal}), in a broad parametric region, provided that the dispersion
coefficient $\beta $ is positive (recall this corresponds to the case of
anomalous chromatic dispersion in the waveguide). As $\beta $ decreases, the
stripes lose their stability against perturbations which rearrange them into
a zigzag pattern. A typical example of a stable zigzag (which coexists with
a stable 2D pulse) is shown in Fig. 7 for $\beta =-0.05,\;\gamma _{1}=2.3$
and $\gamma _{2}=1/2$. Both zig and zag segments of the pattern are nothing
else but pieces of the above-mentioned oblique stripes oriented under equal
but opposite angles relative to the straight stripes. Note that the
transformation (\ref{Galileo}) may as well be applied to the zigzag
solution, generating zigzags with an arbitrary orientation in the ($z,x$)
plane.

\begin{figure}[htb]
\begin{center}
\includegraphics[width=7cm]{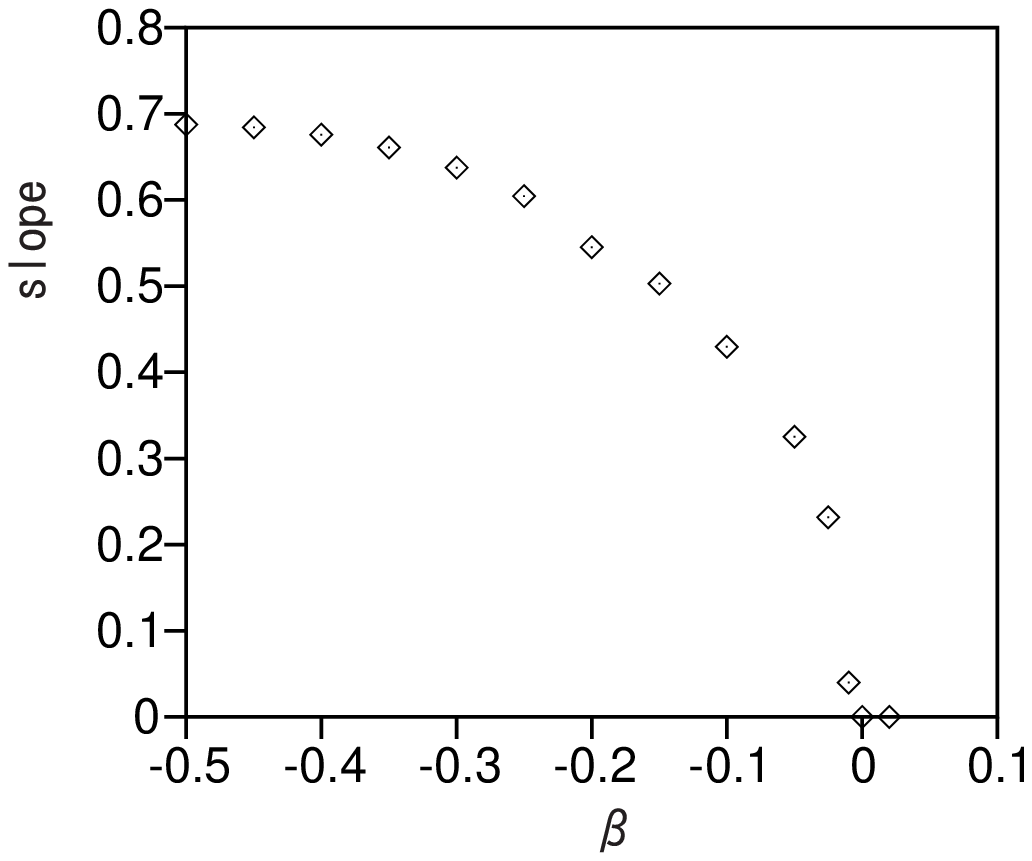}
\caption{An example of the forward bifurcation that destabilizes the
straight-stripe solution and gives rise to a stable zigzag pattern. In this
case, $\gamma _{1}=2.3$ and $\gamma _{2}=1/2$. The absolute value of the
slope of the zigzag's oblique sections is plotted as a function of the
dispersion coefficient $\beta $.} 
\label{fig:8} 
\end{center}
\end{figure} 
While stable straight and zigzag stripes coexist with the stable 2D pulses,
we have found that they never coexist with each other as stable solutions.
Roughly, the straight stripes are stable at $\beta >0$ and unstable at $%
\beta <0$. A bifurcation which destabilizes the straight stripe and gives
rise to the zigzag is a supercritical (forward) one, see a typical example
in Fig. 8. As a parameter characterizing the transition to the zigzag, in
this figure we show the absolute value of the slope of the step's oblique
segments (see Fig. 7) vs. $\beta $ for $L=20$.

\begin{figure}[htb]
\begin{center}
\includegraphics[width=7cm]{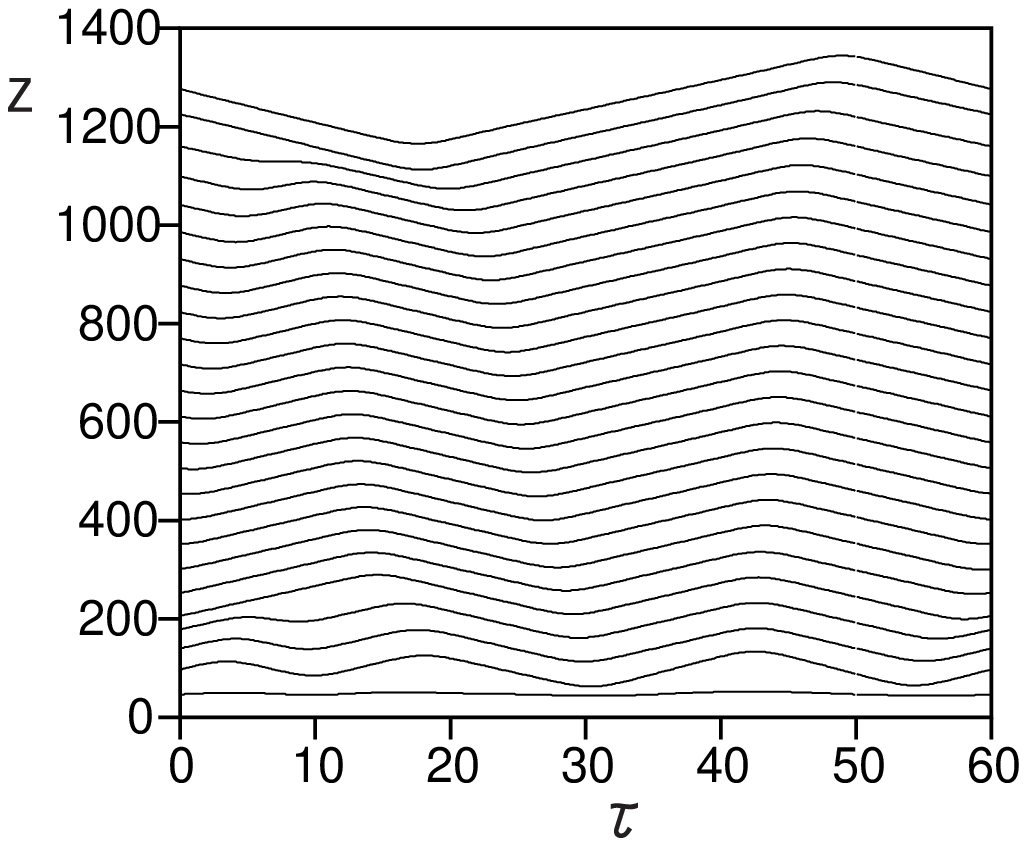}
\caption{Successive shapes of the ridge line $x=x(\tau ,z)$, at which the
field $|u(z,x,\tau )|$ attains its maximum along the $x$ direction, are
shown for an evolving stripe pattern as $z$ increases. The initial state is
a straight unstable stripe with small random perturbations added to it, the
integration domain being expanded in the $\tau $-direction to the size $%
L_{\tau }=60$ (three times the size used in other simulations). At first,
three zigzags appear, which later merge into a single one as a result of the
coarsening process.} 
\label{fig:9} 
\end{center}
\end{figure} 
If the system's size $L_{\tau }$ in the $\tau $ direction is large enough,
the instability of the straight stripe initiated by random perturbations may
at first give rise to multi-zigzag patterns. However, a coarsening process
occurs, and as a result the pattern eventually relaxes to a single-zigzag
state, as is shown in Fig. 9 for $L_{x}=20$ and $L_{\tau }=60$ at $\beta
=-0.05$, $\gamma _{1}=2.3$ and $\gamma _{2}=1/2$. It means that two oblique
stripes with the absolute value of the slope given by the bifurcation
diagram shown in Fig. 8 are stable solutions in the case when the straight
stripe is unstable.

Zigzag patterns are known in models of the complex-Swift-Hohenberg \cite{HS}
and Segel-Newell-Whitehead (SNW) \cite{Hoyle} types; coarsening of the
zigzag structure is also a known feature of the SNW equation. In particular,
in the former model it was found that roll-type standing-wave patterns give
rise to two specific modes of small perturbations around them: one mode
accounts for zigzag corrugation of the stripes, and the other one is a phase
modulation of the oscillations. The two modes are decoupled in the linear
analysis, but they get coupled in a nonlinear regime. The same set of two
modes can be found around the localized straight-stripe pattern in the
present model, hence the onset of the zigzag instability in the present
model is similar to that studied in detail in Refs. \cite{HS}. However,
physical interpretation of each zigzag in terms of the planar optical
waveguide underlying the present model is quite different: a spatial soliton
moving in the transverse direction in the homogeneous planar waveguide
suddenly reverses its velocity to the opposite value. This appears to be the
first example of such a ``boomerang'' motion mode of a stripe soliton in
nonlinear optics (note the GL equation does not conserve momentum).

\section{Conclusion}

In this work, we have introduced a model of a two-dimensional optical
waveguide with the Kerr nonlinearity, linear and quintic losses, cubic gain,
and temporal-domain filtering. In the general case, the model also contains
temporal dispersion, although the latter ingredient is not necessary. The
model may provide for a realistic average description of a nonlinear planar
waveguide incorporated into a closed optical cavity. It takes the form of a
two-dimensional cubic-quintic Ginzburg-Landau equation with an anisotropy of
a novel type: the equation is diffractive in one direction, and diffusive
(or mixed diffusive/dispersive) in the other. Following analogy with the
model's one-dimensional counterpart, we have demonstrated by systematic
direct simulations that the model gives rise to stable fully localized
two-dimensional pulses, which, in terms of the optical planar waveguide, are
spatiotemporal ``light bullets'', existing due to the simultaneous balances
of diffraction, dispersion, and Kerr nonlinearity, and of linear losses
(including those induced by the filtering), cubic gain, and quintic loss. A
stability region of the two-dimensional pulses was identified in the
system's parameter space. We have also found that the model generates stable
(quasi-) one-dimensional patterns in the form of a simple straight stripe or
oblique stripes with zigzags. The straight and oblique stripes may stably
coexist with the two-dimensional pulse, but not with each other. A
bifurcation which destabilizes the straight stripe and gives rise to the
oblique ones was found to be supercritical. In terms of the optical
waveguide, the straight-stripe solution corresponds to a simple spatial
soliton, while the zigzag on oblique stripes may be realized as a spatial
soliton moving in the transverse direction which suddenly changes the sign
of its velocity. It seems to be the first example of such a dynamical
behavior in nonlinear optics.

\section*{Acknowledgements}

The work of B.A.M. was partly supported by a joint grant from the Japan
Society for Promotion of Science and Israeli Ministry of Science and
Technology. This author appreciates hospitality of the Department of
Aeronautics and Astronautics at the Kyoto University, and of the Department
of Applied Physics at the Kyushu University (Fukuoka, Japan).

\end{document}